\documentclass[a4paper,11pt]{article}
\usepackage{jcappub} 
\usepackage{lineno}

\title{\boldmath Impacts of axion cooling on the direct detection of supernova axions}

\author[a,b]{Kanji Mori}
\affiliation[a]{Department of Physics, Faculty of Science and Technology, Keio University, \\3-14-1 Hiyoshi, Kohoku-ku, Yokohama, Kanagawa 223-8522, Japan}
\affiliation[b]{Division of Science, National Astronomical Observatory of Japan, \\2-21-1 Osawa, Mitaka, Tokyo 181-8588, Japan}
\author[c]{and Masamitsu Mori}
\affiliation[c]{National Institute of Technology (KOSEN), Numazu College, \\3600 Ooka, Numazu, Shizuoka 410-0022, Japan}

\emailAdd{kanji.mori.astro@outlook.com}

\abstract{Core-collapse supernovae provide a unique opportunity to probe axions because they can be a copious source of the particles. It has recently been proposed that axion helioscopes can be used for the direct search for supernova axions if a supernova event appears within a few hundred parsecs. However, the event number of supernova axions has been estimated only within the post-process framework. In this study, we perform long-term supernova simulations for a $9.6M_\odot$ star coupled with the axion emission to reevaluate the event number of  axions detected by the helioscopes. We find that the additional cooling induced by the axion emission can significantly decrease the temperature in the proto-neutron star. As a result, the axion luminosity and hence the axion event number are reduced, compared with the result obtained through post-processing. Our result indicates that the nonlinear feedback of the axion emission is an essential factor to predict the axion detectability, and underscores the need for systematic simulation studies across various progenitor models.}

\keywords{dark matter theory, axions, core-collapse supernovae}

\begin{document}

\maketitle
\flushbottom

\section{Introduction}
\label{sec:intro}

The axion is a hypothetical boson which is introduced to solve the strong CP problem in quantum chromodynamics \cite{1978PhRvL..40..223W,1978PhRvL..40..279W}. Since they are also a candidate for dark matter \cite{1983PhLB..120..133A,1983PhLB..120..127P,1983PhLB..120..137D}, massive efforts have been dedicated to experimental and astrophysical searches for axions \cite[e.g.][]{1978JETPL..27..502V,2021ARNPS..71..225C,2024arXiv240113728C,2025PhyR.1117....1C}.

Among astrophysical objects, core-collapse supernovae are one of the most luminous sources of axions because of the high temperature inside. Hence neutrinos and photons from SN~1987A, which appeared in the Large Magellanic Cloud, have been used to constrain the interaction between axions and the Standard Model particles. For example, axions  work as an additional cooling process in the proto-neutron star (PNS), which can render the duration of the neutrino burst shorter than observed \cite{1988PhRvL..60.1797T,PhysRevD.39.1020,1988PhRvL..60.1793R,1990PhRvL..65..960E,1997PhRvD..56.2419K,2016PhRvD..94h5012F,2019JCAP...10..016C,2021PhRvD.104j3012F,PhysRevD.106.063019,2023APh...15102855F,2024PhRvD.109b3001L}. In addition, if axions interact with photons, they can produce observable $\gamma$-rays during the propagation in the interstellar magnetic field \cite{1996PhRvL..77.2372G,1996PhLB..383..439B,2015JCAP...02..006P,2017PhRvD..96e1701K,2024PhRvL.133u1002M}. The axion-photon conversion in the interstellar space may enable one to even detect the signature of the diffuse supernova axion background \cite{2011PhRvD..84j3008R,2020PhRvD.102l3005C,2022PhRvD.105f3028C}. It is also notable that nearby massive stars before the core-collapse can produce a significant amount of axions that are converted into x-rays \cite{1995PhLB..344..245C,2021PhRvL.126c1101X,2022PhRvD.106l3019X,2022PhRvD.105b3020M}.

Apart from SN~1987A and distant stars, the Sun is another useful laboratory to study axions because of its proximity and stationarity. Axion helioscopes including CAST \cite{2017NatPh..13..584A} and Tokyo Axion Helioscope \cite{2008PhLB..668...93I} have observed the Sun to search for Solar axions through the axion-photon conversion in the strong magnetic field. Also, next-generation axion helioscopes such as IAXO \cite{2019JCAP...06..047A} and TASTE \cite{2017JInst..12P1019A}  will lead to tighter constraints on the axion-photon interaction.

Recently, the idea of direct detection of supernova axions with terrestrial detectors has been proposed \cite{2020JCAP...11..059G,2022PhLB..82937137A,2025arXiv250219476C}. In particular, ref.~\cite{2020JCAP...11..059G} proposed using the axion helioscopes. Since supernova axions are typically as energetic as $\sim100$\,MeV, they can be converted to $\gamma$-rays in strong magnetic fields. Therefore, if a $\gamma$-ray detector is installed on the other side of the x-ray detector for Solar axions, the axion helioscopes can be used as \emph{axion supernova-scopes}. Since the axion burst from a supernova lasts only for $\sim10$\,s,  the supernova-scopes should be aimed at the progenitor star before its explosion. Ref.~\cite{2020JCAP...11..059G} proposed to use the presupernova neutrino alert \cite{2004APh....21..303O,2017ApJ...851....6P,2020ARNPS..70..121K,2021NJPh...23c1201A} to target the  supernova-scopes. Although the directional sensitivity of the inverse $\beta$-decay is weak, the unique identification of the progenitor is still possible \cite{2020ApJ...899..153M,2020JCAP...05..049L}.

In refs.~\cite{2020JCAP...11..059G,2025arXiv250219476C}, it is estimated that $\mathcal{O}(1)$ axion events from $\alpha$ Orionis (Betelgeuse) can be detected  by an improved version of IAXO, if the axion-photon coupling constant is as high as the upper limit obtained from SN 1987A neutrinos. However, when the axion luminosity is close to the neutrino luminosity, the proto-neutron star cooling is significantly accelerated \cite{2016PhRvD..94h5012F,2019JCAP...10..016C,2021PhRvD.104j3012F,PhysRevD.106.063019}. As a result, the axion luminosity could be overestimated if one adopts the post-process technique. In this study, we perform long-term supernova simulations coupled with the axion emission to consider the effect of the additional cooling on the supernova axion signal.

This paper is organized as follows. In Section \ref{method}, we describe the setup of our supernova simulations and our benchmark axion model. In Section \ref{results}, we show the results of the simulations and estimate the event number of supernova axions. In Section \ref{conclusion}, we conclude the paper, focusing on the future prospect for more accurate prediction.

\section{Method}
\label{method}
We perform  general-relativistic neutrino radiation hydrodynamics simulations in spherically symmetric geometry by employing the \texttt{GR1D} code \cite{2010CQGra..27k4103O,2015ApJS..219...24O}. The code solves the neutrino transport with the truncated moment formalism \cite{2011PThPh.125.1255S,2013PhRvD..87j3004C}. We use 18 logarithmically spaced energy bins from 1.0 to 280.5\,MeV. We modify and optimize GR1D for long-term simulation \cite{2021PTEP.2021b3E01M,2025PASJ...77..127M}. We utilize the DD2 equation of state \cite{2005PhRvC..71f4301T,2010NuPhA.837..210H}, which is based on the relativistic mean field approach. The neutrino reaction rates are calculated with the \texttt{NuLib} routine \cite{2015ApJS..219...24O}. As the initial condition, we adopt a $9.6M_\odot$ star with zero-metalicity provided by A. Heger (2016, private communication). The progenitor is chosen because the star successfully explodes even in the one-dimensional configuration without enhancing the neutrino heating rate by hand \cite{2015ApJ...801L..24M,2021PTEP.2021b3E01M}. The outer boundary for the simulations is located at the radius of $r_\mathrm{out}=5000$\,km.

The original version of \texttt{GR1D} is optimized to simulate the accretion phase, which is $\lesssim 1$\,s after the core bounce. Since the axion burst can continue after the accretion phase, we perform  simulations until the post-bounce time $t_\mathrm{pb}=20$\,s. In order to realize the long-term simulations, we modify the spatial resolution to resolve the steep density gradient at the surface of the PNS. The number of the grid points is $N_r=300$. In addition, we extend tables for thermodynamic quantities in the code because they can exceed the table in \texttt{GR1D} during the long-term simulations. See ref.~\cite{2021PTEP.2021b3E01M} for details on the modification.

We calculate the axion emissivity during the simulations. We consider the Kim-Shifman-Vainshtein-Zakharov (KSVZ) axions \cite{1979PhRvL..43..103K,1980NuPhB.166..493S} as the benchmark model. Axions can interact with nucleons and photons as described by the Lagrangian \cite{1989PhRvD..40..652C}
\begin{equation}
    \mathcal{L}=\sum_{N=p,\;n}\frac{g_{aN}}{2m_N}\bar{N}\gamma^\mu\gamma_5N\partial_\mu a-\frac{g_{a\gamma}}{4}aF_{\mu\nu}\tilde{F}_{\mu\nu},
\end{equation}
where $g_{aN}$ is the axion-nucleon coupling, $g_{a\gamma}$ is the axion-photon coupling, $N$ is nucleons, $a$ is axions, $F_{\mu\nu}$ is the electromagnetic tensor and $\tilde{F}_{\mu\nu}$ is its dual, and $m_\mathrm{N}$ is the nucleon mass. The coupling constants for KSVZ axions are given as
$g_{aN}=C_{aN}m_N/f_a$ and $g_{a\gamma}=\alpha/\pi f_a$, 
where $m_a$ is the axion mass, $f_a$ is the axion decay constant, $C_{ap}=-0.47(3)$, and $C_{an}=-0.02(3)$ \cite{2016JHEP...01..034D}. The axion mass and the decay constant have the relationship 
\begin{equation}
    m_a=\frac{\sqrt{m_um_d}}{m_u+m_d}\frac{f_\pi m_\pi}{f_a},
\end{equation}
where $m_u$, $m_d$, $m_\pi$ are the mass of up quarks, down quarks, and neutral pions, respectively, and $f_\pi$ is the pion decay constant. We adopt the mass range of $m_a=3$--11\,meV in our simulations, considering the SN~1987A upper limit $m_a\lesssim5$--15\,meV \cite{2019JCAP...10..016C}. Table \ref{tab:1} is the list of our models. As shown in the table, we perform six simulations with axions and a simulation without axions.

 \begin{table}[tbp]
 \centering
 \begin{tabular}{c|ccccccc}
 \hline
 $m_a$&$f_a$ &$|g_{ap}|$  &$L_a$ &$N$ (sim.)&$N$ (post-process) & Ratio \\
 $[\mathrm{meV}]$&[$10^8$\,GeV]&[$10^{-10}$]&[$10^{52}$\,erg\,s$^{-1}$]&&&\\
 \hline
 (NoAxion) & -- & --& 0&0&0&--\\
 3 & 19 & 2.3 & 0.3 & 0.0087 & 0.014 &0.64\\
 6&9.7&4.7 & 1.3 & 0.074 &0.22 &0.34 \\
 8&7.3&6.2&2.3&0.16&0.69&0.24\\
 9&6.4&7.0&2.9&0.23&1.10&0.21\\
 10&5.8&7.8&3.5&0.30&1.68&0.18\\
 11&5.3&8.5&4.3&0.39&2.45&0.16\\
 \hline
 \end{tabular}
 \caption{The summary for our models. The first column shows the axion mass, the second shows the decay constant and the third shows the axion-proton coupling constant. The fourth column shows the axion luminosity measured at the post-bounce time $t_\mathrm{pb}=1$\,s. The fifth and sixth columns show the axion event number from Betelgeuse detected by upgraded IAXO based on the self-consistent simulations and post-processing of the reference model. The seventh column shows the ratio between the event numbers from simulations and the post-processing. \label{tab:1}}
 \end{table}

The dominant production channel of supernova axions is the nucleon-nucleon bremsstrahlung
\begin{equation}
    NN'\rightarrow NN'+a,
\end{equation}
where $N$ and $N'$ are nucleons. Within the one-pion exchange (OPE) approximation, the axion emissivity in vacuum is given as \cite{1988PhRvD..38.2338B,1995PhRvD..52.1780R,2019JCAP...10..016C}
\begin{eqnarray}
    Q_a^\mathrm{vac}=64\left(\frac{f}{m_\pi}\right)m_N^\frac{5}{2}T^\frac{13}{2}\left(\left(1-\frac{\bar{\xi}}{3}\right)g_{an}^2I(y_n,\,y_n)+\left(1-\frac{\bar{\xi}}{3}\right)g_{an}^2I(y_p,\,y_p)\right.\nonumber\\
    \left.+\frac{4(3+2\bar{\xi})}{9}\left(\frac{g_{an}^2+g_{ap}^2}{2}\right)I(y_n,\,y_p)+\frac{8(3+2\bar{\xi})}{9}\left(\frac{g_{an}+g_{ap}}{2}\right)^2I(y_n\,y_p)\right),
\end{eqnarray}
where $f\approx1$ is the pion-nucleon fine structure constant \cite{1998ApJ...507..339H}, $T$ is the temperature, $\bar\xi$ is a parameter to represent the degeneracy \cite{1995PhRvD..52.1780R}, $y_n$ and $y_p$ are the nucleon degeneracy parameters, and the function $I(y_1,\,y_2)$ is a fitting formula developed in ref.~\cite{1988PhRvD..38.2338B}. Taking into account the many-body effect that becomes important in the high density matter, we adopt in the simulations the axion emission rate \cite{1997PhRvD..56.2419K,2016PhRvD..94h5012F}
\begin{equation}
    Q_a=Q_a^\mathrm{vac}\min\left(\frac{\Gamma_\sigma^\mathrm{max}}{\Gamma_\sigma},\,1\right),
\end{equation}
where 
\begin{equation}
    \Gamma_\sigma\approx10\,\mathrm{MeV}\,\left(\frac{\rho}{10^{14}\,\mathrm{g\,cm^{-3}}}\right)\left(\frac{T}{1\,\,\mathrm{MeV}}\right)^\frac{1}{2}
\end{equation}
is the lowest-order spin fluctuation rate, $\rho$ is the density, and $\Gamma_\sigma^\mathrm{max}=60$\,MeV. The internal energy $\epsilon$ per volume at the $(n+1)$-th step is then estimated as $\epsilon^{n+1}=\epsilon^n-Q_a\Delta t$, where $\Delta t$ is the time step.

The OPE treatment for the nucleon bremsstrahlung has been often adopted in literature to estimate the axion emission from supernovae \cite[e.g.][]{1997PhRvD..56.2419K,2016PhRvD..94h5012F,PhysRevD.106.063019}. However, treatments beyond the OPE are recently discussed \cite[e.g.][]{2018JHEP...09..051C,2019JCAP...10..016C}. In ref.~\cite{2019JCAP...10..016C}, the effects of the nonzero pion mass \cite{2009NuPhA.828..439S}, the two-pions exchange \cite{1989PhLB..219..507E}, the effective nucleon mass, and the nucleon multiple scatterings \cite{1991PhRvL..67.2605R,1996PhRvL..76.2621J} are considered to calculate the axion emission from a supernova. They found that the axion luminosity is suppressed by an order of magnitude if these corrections are taken into account. In addition, it has been pointed out that supernova axions can be produced through other channels such as the pion-induced reaction ($\pi^- p\rightarrow na$) \cite{1995PhRvD..52.1780R,1997PhRvD..56.2419K,2021PhRvL.126g1102C,2021PhRvD.104j3012F} and a large number of reactions involved with strange matter \cite{2024PhRvL.133l1002C}. These additional channels will enhance the axion emissivity. In particular, refs.~\cite{2021PhRvL.126g1102C,2021PhRvD.104j3012F} discuss that the pion-induced reactions produce high-energy axions, which could be detected by future water-Cherenkov detectors through the axion-nucleon interaction.  In order to predict the axion luminosity from supernovae accurately, it is desirable to perform simulations that take into account these corrections beyond the OPE and the additional axion production channels. However, since our motivation here is to investigate the importance of the axion cooling effect on the axion detectability, we leave this task for future work.

\section{Results}
\label{results}

The core-collapse of the star produces a PNS with  the baryonic mass $\sim1.36M_\odot$ at the center. Since the emission of axions works as a new cooling channel, the thermal evolution of the PNS will be affected by axions. The interplay between the PNS evolution and the axion emission can affect the feasibility of the direct detection of supernova axions. In this Section, we describe the effects of axion cooling on the PNS evolution on the basis of our simulations.

\subsection{Axion and neutrino luminosities}

 \begin{figure}[tbp]
 \centering
 \includegraphics[width=.9\textwidth]{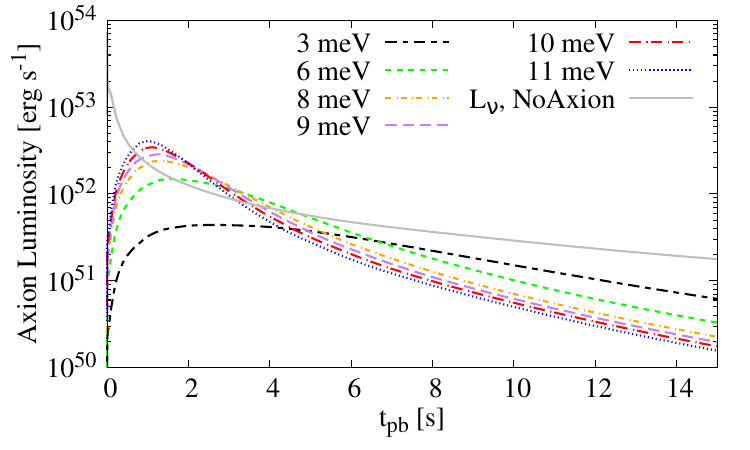}

 \caption{The axion luminosity for each model is shown in the broken curves. The solid curve is the neutrino luminosity for the reference model without axions. \label{fig:axionL}}
 \end{figure}

  \begin{figure}[tbp]
 \centering
 \includegraphics[width=.45\textwidth]{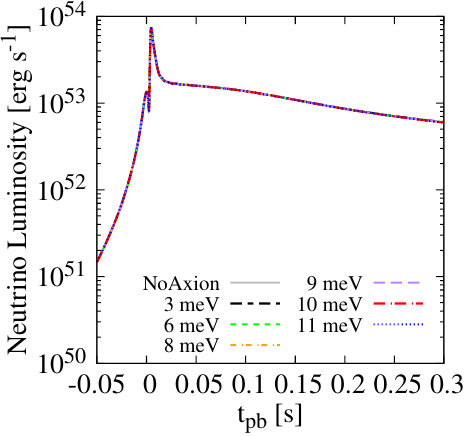}
 \qquad
 \includegraphics[width=.44\textwidth]{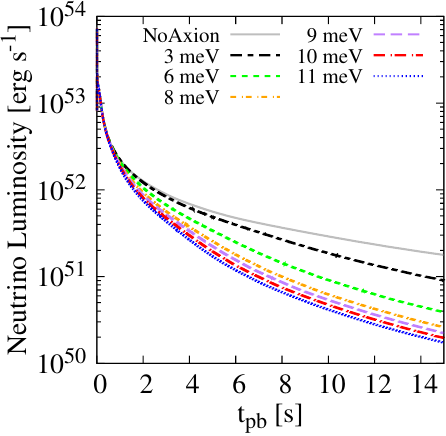}

 \caption{The neutrino luminosity for each model. The left panel focuses on the early phase at $t_\mathrm{pb}\in[-0.05,\,0.3]$\,s and the right panel shows the late phase at $t_\mathrm{pb}\in[0,\,15]$\,s. The solid curve represents the result for the reference model without axions and the broken curves represent the results with axions.\label{fig:neutrinoL}}
 \end{figure}

  \begin{figure}[htbp]
 \centering
 \includegraphics[width=.42\textwidth]{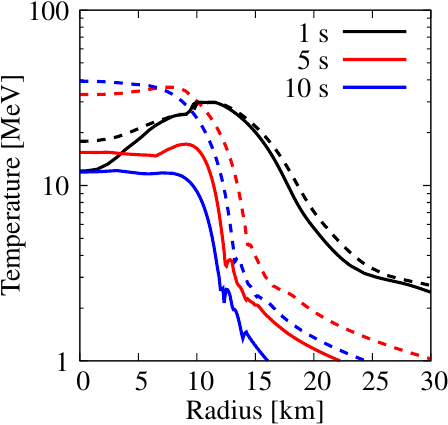}
 \qquad
 \includegraphics[width=.45\textwidth]{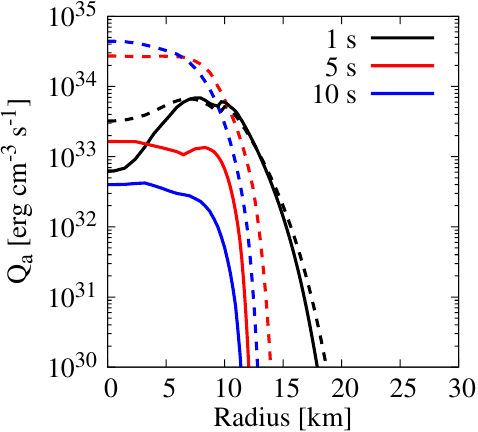}
 \caption{The temperature (left panel) and $Q_\mathrm{a}$ (right panel) profiles. The solid curves are the results for the $m_a=10$\,meV model and the broken curves are for the reference model without the axion cooling effect. In the right panel, $Q_a$ for the reference model is estimated with the post-processing technique, assuming $m_a=10$\,meV. \label{fig:temp}}
 \end{figure}
 
Figure \ref{fig:axionL} shows the axion luminosity for each model, which is estimated as
\begin{eqnarray}
L_a=4\pi\int_0^{r_\mathrm{out}} Q_a r^2dr,    
\end{eqnarray}
where $r_\mathrm{out}$ is the outer boundary of the simulations. One can find that the peak is achieved at the post-bounce time of $t_\mathrm{pb}\sim1$--2\,s. After that, the axion luminosity gradually decreases. The peak luminosity is higher when axions are heavier because the axion-proton coupling constant is proportional to the axion mass. However, the luminosity decreases more slowly when axions are lighter. As a result, at the late phase at $t_\mathrm{pb}\gtrsim6$\,s, lighter axions lead to higher axion luminosities, although the coupling constant is smaller.

Figure \ref{fig:neutrinoL} shows the neutrino luminosity for the models. In this figure, the contributions of all the neutrino flavors are summed up. As the PNS is cooled down, the neutrino luminosity decreases. Since axions carry away the energy from the PNS, the decay of the neutrino luminosity becomes faster when axions are considered. The neutrino luminosity is smaller when axions are heavier because the total energy of emitted axions is larger.

In figure \ref{fig:axionL}, the neutrino luminosity for the reference model without axions is also shown. One can find that, when $m_a\geq6$\,meV, the axion peak luminosity exceeds the neutrino luminosity. However, the axion luminosity decays faster than the neutrino luminosity, and it becomes subdominant over the neutrino luminosity at $t_\mathrm{pb}> 3$--5\,s. On the other hand, in the $m_a=3$\,meV model, which shows the lowest axion peak luminosity  among our models, the axion luminosity is always lower than the neutrino luminosity. 

Since $Q_a$ is approximately proportional to $g_{ap}^2$, one can expect that the axion luminosity is proportional to the axion mass. However, as shown in figure \ref{fig:axionL}, the axion light curve does not follow the simple relation. This behavior can be attributed to the additional PNS cooling induced by the axion emission. The left panel in figure \ref{fig:temp} shows the temperature profile in the supernova core at $t_\mathrm{pb}=1$, 5, and 10\,s for the reference model and the $m_a=10$\, meV model. It can be seen that the temperature is lower when axions are considered. For example, at $t_\mathrm{pb}=10$\,s, the central temperature is as high as $\sim30$\,MeV, while it is $\sim12$\,MeV for the $m_a=10$\,meV model. This difference in the temperature is caused by the energy loss induced by the axion emission. Since $Q_a$ is sensitive to the temperature, the axion cooling affects the axion emissivity as well. The right panel in figure \ref{fig:temp} shows the $Q_a$ profile with $m_a=10$\,meV. In this panel, the solid lines indicate the results from the self-consistent simulation and the broken lines indicate the results obtained by the post-processing. One can find that $Q_a$ is significantly reduced when the axion cooling is considered. At $t_\mathrm{pb}=10$\,s, $Q_a$ at the center is two orders of magnitudes smaller  because the central temperature is reduced. This reduction in $Q_a$ results in the accelerated decay of the axion light curve, which is shown in figure \ref{fig:axionL}.

\subsection{Expected event number of axions}

Axions produced by a nearby supernova event could be detected by axion helioscopes. In the following part, we estimate the axion event number on the basis of our simulations.

An axion helioscope \cite{1983PhRvL..51.1415S} is an instrument to detect Solar axions. A strong magnetic field is imposed in a cylindrical hollow to convert axions into x-rays through the axion-photon coupling. An x-ray detector is installed at the end of the cylinder to detect the signal. Axion helioscopes have been used to obtain tight upper limits on the axion-photon coupling $g_{a\gamma}$. For example, the CAST experiment reported an upper limit $g_{a\gamma}<0.66\times10^{-10}$\,GeV$^{-1}$ at 95\% confidence level \cite{2017NatPh..13..584A}.

As proposed in ref.~\cite{2020JCAP...11..059G}, axion helioscopes can be used to detect supernova axions in principle. One of differences between Solar axions and supernova axions is their energy. Since supernova axions are as energetic as 10--100\,MeV, they are converted into $\gamma$-rays by the magnetic field. In order to detect supernova axions with axion helioscopes, one should install an additional $\gamma$-ray detector. Another difference is that, whereas the Sun is a static object, supernova explosions are transient events. It is thus necessary that the axion helioscope is pointed to the supernova event before its neutrino burst is observed. This requires the identification of the progenitor star before the explosion, which could be possible through a presupernova neutrino alert \cite{2004APh....21..303O,2017ApJ...851....6P,2020ARNPS..70..121K,2020ApJ...899..153M,2020JCAP...05..049L,2021NJPh...23c1201A}. It is estimated that presupernova neutrinos can provide the alert $\mathcal{O}(1)$ hours before the core bounce for a massive star closer than 1\,kpc, with detectors including the Jiangmen Underground Neutrino Observatory (JUNO) \cite{2015arXiv150807166A,2020JCAP...05..049L}, Super-Kamiokande \cite{2003NIMPA.501..418F,2022ApJ...935...40M,2024ApJ...973..140A}, Hyper-Kamiokande \cite{2018arXiv180504163H}, KamLAND \cite{2014EPJC...74.3094S,2016ApJ...818...91A,2024ApJ...973..140A}, the Deep Underground Neutrino Experiment (DUNE) \cite{2020JInst..15.8008A}, and SNO+ \cite{2021arXiv210411687S}. 

The axion-photon conversion probability in an axion helioscope can estimated as \cite{1983PhRvL..51.1415S,1988PhRvD..37.1237R}
\begin{eqnarray}
   P=\frac{1}{4}(g_{a\gamma}BL)^2\left(\frac{\sin\frac{qL}{2}}{\frac{qL}{2}}\right)^2, 
\end{eqnarray}
where $B$ is the magnetic field, $L$ is the length of the helioscope,  $q=|m_a^2-m_\gamma^2|/2\omega$, $m_\gamma$ is the plasma frequency, and $\omega$ is the axion energy. The  helioscope can  probe axions efficiently if $qL\ll1$. In particular, if the axion helioscope is evacuated, $m_\gamma=0$ and thus the condition for the efficient conversion is $m_a\ll\sqrt{2\omega/L}$. For such light axions, the conversion probability is simplified as $P=(g_{a\gamma}BL)^2/4$. 

 \begin{figure}[tbp]
 \centering
 \includegraphics[width=.9\textwidth]{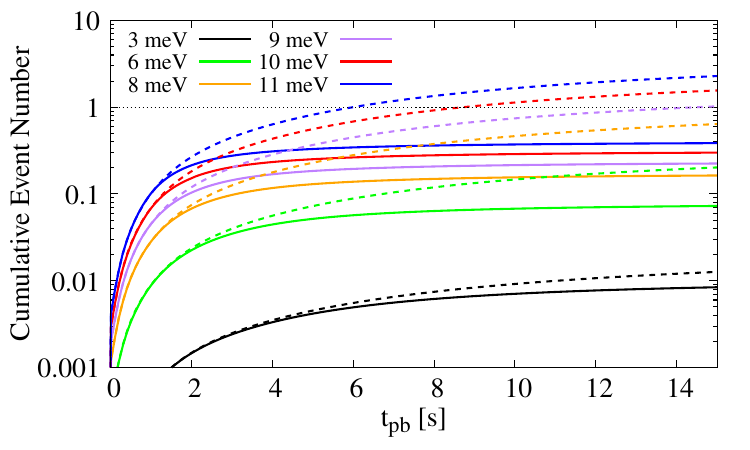}

 \caption{The cumulative event number for axions observed with the IAXO upgraded scenario. The supernova event of Betelgeuse, which is located at $d\approx168$\,pc, is assumed. The solid curves show the result with the axion cooling effect and the broken curves are the results without the effect.\label{fig:eventnum}}
 \end{figure}

  \begin{figure}[tbp]
 \centering
 \includegraphics[width=.42\textwidth]{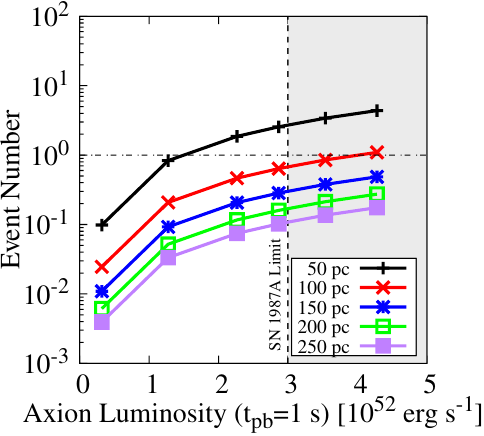}
  \qquad
 \includegraphics[width=.42\textwidth]{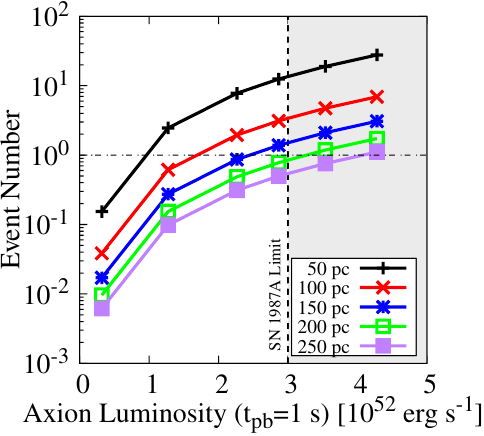}
 \caption{The asymptotic event number of axions as a function of the axion luminosity $L_a$ at $t_\mathrm{pb}=1$\,s. The left panel is based on the self-consistent simulations and the right panel is based on the post-processing. Each curve represents the results with different distances to the supernova event. The IAXO upgraded scenario is assumed as the detector. The vertical line at $L_a=3\times10^{52}$\,erg\,s$^{-1}$ is an upper limit on the axion luminosity based on the SN~1987A neutrino burst \cite{1990PhR...198....1R}. \label{fig:limit}}
 \end{figure}

The axion event number can be estimated as
\begin{eqnarray}
    N=\frac{A}{4\pi d^2}\int P\dot{N}_adt,
\end{eqnarray}
where $A$ is the cross section of the helioscope, $d$ is the distance to the supernova event, $\dot{N}_a=L_a/\langle\omega\rangle$ is the axion number produced in the supernova per unit time, and $\langle\omega\rangle\approx2.3T$ is the axion mean energy. The cross section $A$, the magnetic field $B$, and the length $L$ depend on the detectors. In this study, we adopt values for the \textit{IAXO upgraded} scenario shown in ref.~\cite{2019JCAP...06..047A}, where $B\sim 35$\,kG, $L=22$\,m, and $A=3.9$\,m$^2$. 

Figure \ref{fig:eventnum} shows the cumulative event number of axions from Betelgeuse, which is located at $d\approx168$\,pc \cite{2020ApJ...902...63J}. The solid curves are the results for the models with the axion cooling effect, and the broken curves are the results obtained by the post-processing of the reference model without the axion cooling. One can find that, when the axion cooling is not considered, the event number exceeds one if $m_a\gtrsim9$\,meV. On the other hand, when the axion cooling is taken into account, the cumulative event number saturates earlier and its asymptotic value is significantly reduced. As a result, the expected event number does not exceed one even if axions are as heavy as $m_a=11$\,meV. The reduction in the event number stems from the reduction in the axion luminosity, which is attributed to the lower temperature in the supernova core shown in figure \ref{fig:temp}.

 \begin{table}[tbp]
 \centering
 \begin{tabular}{lcccc}
 \hline
 Catalog Name & Common Name & Spectral Type &Distance [pc] & Mass [$M_\odot$]\\
 \hline
 HD~116658 &Spica/$\alpha$ Virginis &B&77(4) \cite{2007AA...474..653V}&$11.43\pm1.15$ \cite{2016MNRAS.458.1964T}\\
 HD~149757 & $\zeta$ Ophiuchi &O& 112(2) \cite{2007AA...474..653V} & 20.0 \cite{2001MNRAS.327..353H}\\
 HD~129056 & $\alpha$ Lupi & B &143(3) \cite{2007AA...474..653V} & $10.1\pm1.0$ \cite{2011MNRAS.410..190T}\\
 HD~39801 & Betelgeuse/$\alpha$ Orionis & M & $168^{+27}_{-15}$ \cite{2020ApJ...902...63J} & 16.5-19 \cite{2020ApJ...902...63J}  \\
 HD~148478 & Antares/$\alpha$ Scorpii & M & 169(30) \cite{2007AA...474..653V}&  11.0-14.3 \cite{2013AA...555A..24O} \\
 \hline
 \end{tabular}
 \caption{The list of nearby stars heavier than $10M_\odot$ and closer than 200\,pc \cite{2020ApJ...899..153M}.  \label{tab:2}}
 \end{table}

Table \ref{tab:1} shows the asymptotic event number of axions from Betelgeuse for each model. It is estimated that the axion cooling reduces the event number by $\sim40$--$80\%$, depending on the axion mass. Since heavier axions lead to a higher axion luminosity, the reduction is more significant for heavier axions. 

Figure \ref{fig:limit} shows the asymptotic event number of axions as a function of the axion luminosity at $t_\mathrm{pb}=1$\,s. As expected, the axion event number increases when the axion luminosity is higher. In the post-process framework, the axion event number is proportional to $L_a^2$. This is because  $L_a$ is approximately proportional to $m_a^2$, the axion-photon conversion probability $P$ is proportional to $m_a^2$, and thus the event number is proportional to $m_a^4$. On the other hand, this relation does not hold when the axion cooling is considered because of its nonlinear feedback on the axion luminosity.

The axion luminosity has an upper limit based on the neutrino burst from SN~1987A. If the axion luminosity dominates over the neutrino luminosity, the duration of the SN~1987A neutrino burst should be shorter than observed. This energy-loss argument has been applied to various exotic particle models to obtain constraints on the interaction between the new particle and a Standard Model particle. In many cases, the criterion of $L_a<3\times10^{52}$\,erg\,s$^{-1}$ at $t_\mathrm{pb}=1$\,s is adopted for this purpose \cite[e.g.][]{1990PhR...198....1R,2011JCAP...01..015G,2020JCAP...12..008L}. The vertical line in figure \ref{fig:limit} indicates the SN~1987A limit. In our case, the models with $m_a\leq9$\,meV satisfies the condition, regardless of whether the axion cooling is considered or not.  If we respect this upper limit on the axion luminosity, in the post-process framework, the axion signal is detectable if the star is closer than $d\sim150$\,pc. On the other hand, if we consider the axion cooling, the star should be closer than $d\sim100$\,pc for us to detect axions from it.

Since massive stars are a minority in the stellar population, the number of nearby massive stars is limited. Table \ref{tab:2} is a list of stars heavier than $10M_\odot$ and closer than $d=200$\,pc \cite{2016MNRAS.461.3296N,2020ApJ...899..153M}. Among these stars, Betelgeuse and $\alpha$ Scorpii (Antares) are classified spectroscopically as M-type stars, which correspond to red supergiants. Although  their evolutionary stages are under debate \cite[e.g.][]{2020ApJ...902...63J,2022MNRAS.516..693N,2022ApJ...927..115L,2023MNRAS.526.2765S}, they could cause the core-collapse in near future. On the other hand, $\alpha$ Virginis (Spica) and $\alpha$~Lupi are B-type stars and $\zeta$ Ophiuchi is an O-type star. These stars are expected to be younger than red supergiants, and recent stellar models for Spica \cite{2016MNRAS.458.1964T} and $\zeta$ Ophiuchi \cite{2021ApJ...923..277R} suggest that their evolutionary stages do not reach the helium-burning phase, which is $\sim$Myr before the core-collapse. 

Comparing figure \ref{fig:limit} and table \ref{tab:2}, we can see that the nearby massive stars in the table can be a promising target for the direct search of supernova axions if the post-processing is adopted. However, when the feedback of the axion cooling is considered, the expected event number for the stars except for Spica becomes lower than unity. Since our supernova model and the formulation for the axion production rate are still subject to large systematic uncertainties, the feasibility of the direct detection of axions from each star is highly uncertain. Nevertheless, this result indicates that the axion cooling effect is a critical factor for the prediction of axion detectability.

We can obtain a condition on the axion helioscope for direct detection of supernova axions from Betelgeuse that satisfy the SN~1987A limit. Since the axion event number is proportional to $AB^2L^2$, the condition becomes
\begin{eqnarray}
    \left(\frac{A}{3.9\,\mathrm{m}^2}\right)\left(\frac{B}{35\,\mathrm{kG}}\right)^2\left(\frac{L}{22\,\mathrm{m}}\right)^2\gtrsim\frac{1}{0.23}\approx4.3,
\end{eqnarray}
where each quantity is normalized by the values for the IAXO upgraded scenario. The number in the right hand side, 0.23, originates from the expected  event number for the $m_a=9$\,meV model, which shows the highest axion luminosity that satisfies the SN~1987A limit.

\subsection{Total energy carried away by neutrinos and axions}

  \begin{figure}[tbp]
 \centering
 \includegraphics[width=.9\textwidth]{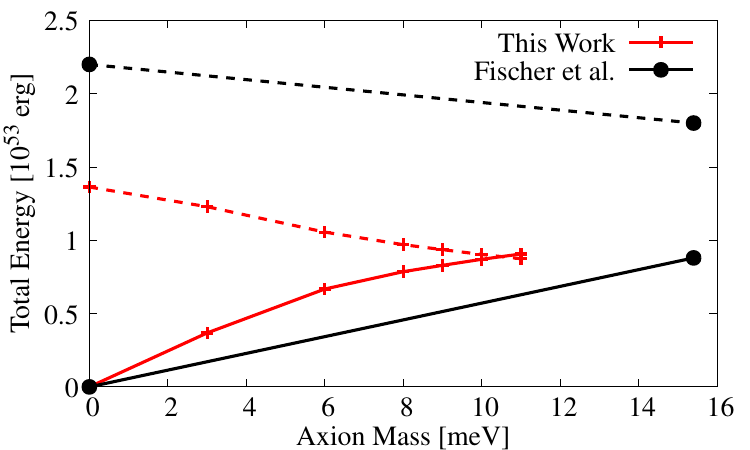}
  \qquad

 \caption{The total energy carried away by axions and neutrinos. The axion total energy is shown in the solid curves and the neutrino total energy is shown in the broken lines. The red curves represent our results and the black curves represent the results from the $aNN$ model in ref.~\cite{2021PhRvD.104j3012F}. \label{fig:TotE}}
 \end{figure}

Since neutrinos and axions can escape from the gravitational well of the star, they carry away the PNS binding energy. In figure \ref{fig:TotE}, the solid curves show the total energy carried away by axions, $E_a$, and the broken curves show the energy carried away by neutrinos, $E_\nu$. We can see that $E_a$ increases when axions are heavier because of the higher axion luminosity. On the other hand, $E_\nu$ decreases as a function of $m_a$ because the neutrino luminosity is reduced by the additional PNS cooling through the axion emission. We also note that, in the post-processing framework,  $E_a$ is expected to be proportional to $m_a^2$. However, figure \ref{fig:TotE} shows that the increase in $E_a$ is slower than $m_a^2$ because the axion luminosity is suppressed by axion cooling itself.

Long-term simulations for PNS cooling coupled with the axion emission have been performed in refs.~\cite[e.g.][]{PhysRevD.39.1020,1997PhRvD..56.2419K,2016PhRvD..94h5012F,2019JCAP...10..016C,2021PhRvD.104j3012F}. In figure \ref{fig:TotE}, we show the result from a recent simulation in ref.~\cite{2021PhRvD.104j3012F}. One can find that, as axions carry away the PNS binding energy, the total neutrino energy becomes suppressed. It is notable that, whereas the simulation setups such as the numerical codes and the progenitor models are different for each model, the difference of $E_a$ falls within a factor of two. 

\section{Conclusion}
\label{conclusion}

In this study, we performed long-term neutrino radiation hydrodynamic simulations for supernova explosion with the axion cooling effect. Based on our models, we estimated the event number of axions from a nearby supernova event detected by axion helioscopes. We found that the axion cooling reduces the axion luminosity when it is comparable to the neutrino luminosity, and as a result, the detected event number becomes smaller than the result obtained by the post-processing technique. This result indicates that one should consider the feedback effect of the axion cooling to evaluate accurately the feasibility of the direct detection of supernova axions.

 In our simulations, the progenitor mass is fixed to $9.6M_\odot$ because the star explodes  without artificial treatments even in the spherically-symmetric geometry. However, the axion luminosity can be dependent on the progenitor models because it is sensitive to the core temperature, as reported in refs.~\cite{2022PhRvD.105f3028C,2025arXiv250309005T}. In order to obtain successful explosions for heavier stars, one should perform multi-dimensional simulations. Recently, axisymmetric supernova simulations coupled with axions (or axion-like particles) have been performed \cite{PhysRevD.106.063019,2023PhRvD.108f3027M}, but these models are limited to the accretion phase at $t_\mathrm{pb}\lesssim1$\,s.  It is desirable to perform long-term multi-dimensional simulations coupled with the axion cooling for a wide range of progenitors to investigate the detectability of supernova axions from each star.

It should be noted that current supernova models suffer from significant uncertainties in microphysics \cite[e.g.][]{Kotake06,Janka12,Burrows21,Yamada2024}. For example, our simulations do not consider neutrino collective oscillations, which can have qualitative impacts on the supernova explosion mechanism \cite[e.g.][]{2019PhRvD.100d3004A,2019ApJ...886..139N,2021PhRvD.103f3033A,2021PhRvD.104h3025N,2023PhRvD.108l3024L,2023PhRvD.107h3016X,Nagakura2023,Ehring2023a,Ehring2023b,2024PhRvD.109b3012A,2025PASJ...77...L9M,2025arXiv250722985F}. Also, the effects of different nuclear equations of state \cite[e.g.][]{Sumiyoshi2020,2021PrPNP.12003879B} and neutrino transport schemes \cite[e.g.][]{2009ApJ...698.1174L,2011PThPh.125.1255S,2012ApJS..199...17S,2013PhRvD..87j3004C} should be investigated. Better understanding of the standard neutrino-heating explosion mechanism will lead to a more accurate prediction of the axion emission from supernova events. At the same time, as discussed in Section \ref{method},  treatments of the axion production rates require further investigations.

\acknowledgments

The authors thank Yudai Suwa and Tomoya Takiwaki for discussions. Numerical analyses were in part carried out on analysis servers at the Center for Computational Astrophysics, National Astronomical Observatory of Japan. This work is supported by JSPS KAKENHI Grant Numbers  JP23KJ2147, JP23K13107, JP23KJ2150, and JP25H02194.




\bibliography{biblio}
\bibliographystyle{JHEP}

\end{document}